\documentclass[aps,prb,citeautoscript,twocolumn,floatfix,showpacs]{revtex4} 
\usepackage{graphicx}
\usepackage{amsmath}
\usepackage{subfigure}
\usepackage{dcolumn}
\usepackage{bm}

\begin{document}

\title{
Pfaffian pairing and backflow wave functions for
electronic structure\\ quantum Monte Carlo methods}

\author{M. Bajdich}\email{mbajdic@ncsu.edu}
\author{L. Mitas} \author{L. K. Wagner}
\affiliation{
Center for High Performance Simulation and Department of Physics,
North Carolina State University, Raleigh, NC 27695}

\author{K. E. Schmidt}
\affiliation{Department of Physics,
Arizona State University, Tempe, AZ 85287}

\date{\today}

\begin{abstract}
We investigate pfaffian trial wave functions with singlet and triplet pair 
orbitals by quantum Monte Carlo methods.
We present mathematical identities and the key algebraic properties 
necessary for efficient evaluation of pfaffians.
Following upon our previous study~\cite{pfaffianprl}, we 
explore the possibilities of expanding the wave function in linear combinations of pfaffians. 
We observe that molecular systems require much larger expansions 
than atomic systems and linear combinations of a few pfaffians 
lead to rather small gains in correlation energy.
We also test the wave function based on fully-antisymmetrized product
of independent pair orbitals. Despite its seemingly large variational potential,
we do not observe additional gains in correlation energy. 
We find that pfaffians lead to substantial improvements in fermion nodes 
when compared to Hartree-Fock wave functions and exhibit the minimal number
of two nodal domains in agreement with recent results on fermion nodes topology. 
We analyze the nodal structure differences of Hartree-Fock, pfaffian and
essentially exact large-scale configuration interaction wave functions. 
Finally, we combine the recently proposed form of backflow correlations~\cite{drummond_bf,rios_bf} 
with both determinantal and pfaffian based wave functions.
\end{abstract}

\maketitle
 
\section{\label{sec:level1} Introduction}
One of the most promising 
many-body electronic structure approaches 
is the quantum Monte Carlo (QMC) method, which employs stochastic techniques for
solving the stationary Schr\" odinger equation
and for evaluation of expectation values~\cite{qmchistory1,qmchistory2,hammond,qmcrev}.
QMC methodology has an important virtue that it enables us
to test and employ variety of many-body wave functions (WFs) with explicit electron--electron
correlation. This opens a possibility to explore wave functions,
which are very difficult to use with traditional methods
based on one-particle basis expansions and on orthogonality of one-particle
orbitals. These high accuracy wave functions enable us to understand 
the nature of many-body effects and also to decrease the QMC fixed-node
errors which come from the necessity to circumvent the fermion sign 
problem~\cite{jbanderson75,reynolds82}.
Fixed-node QMC has been very  effective in providing high accuracy results
for many real systems such as molecules, clusters 
and solids with  
hundreds of valence electrons. 
Typically, for cohesive and binding energies, band gaps, 
and other energy differences the agreement with experiments is within 1-3\% \cite{qmcrev,jeff_benchmark}.
The key challenge for successful application of fixed-node QMC is to develop methods, 
which can eliminate the fixed-node bias or at least make it smaller 
than experimental error bars for the given quantity.
This is a difficult task, once we realize that
the fermion nodes (the subset of position space
where the wave function vanishes)
are complicated high-dimensional manifolds determined 
by the many-body effects. So far, improvement in the accuracy of trial wave functions 
has proved to be one realistic approach to finding better approximations 
for the nodes. This approach has an additional benefit in forcing us to 
think about the relevant correlation effects and their 
compact and computationally efficient description.
 
The commonly used  QMC trial wave functions have          
the Slater--Jastrow  form, which  
can be written as  $ \Psi_{T}=\Psi_{A} \exp[U_{corr}]$
where $\Psi_A$ is the antisymmetric part while $U_{corr}$
describes the electron-electron and higher-order correlations.
The antisymmetric component is typically one or a linear
combination of several Slater determinants of one-particle orbitals,
such as Configuration Interaction (CI) expansion~\cite{szabo}.
To overcome the limit of one-particle orbitals, the two-particle
or pair orbital has been suggested.  In condensed systems, one 
such example is the Bardeen--Cooper--Schrieffer (BCS) wave function\cite{BCS}, 
which is an antisymmetrized product of identical singlet pairs. 

The pair orbital is also referred to as geminal and the resulting wave 
function as the antisymmetrized product of geminals (APG)\cite{fockAPG,HurleyAPG}.  
In its full variational limit, say, for a system with $M$ pairs of fermions, 
such a wave function is an antisymmetrized product of $M$ distinct pair orbitals.
However this freedom leads to computationally very demanding form, since the number of terms 
grows exponentially with the number of pairs. In this work, we have actually 
tested the APG wave functions on a few-particle systems as will be reported later.
On the other hand, if all the geminals in the product are identical, 
then this special case of APG is called the antisymmetrized geminal power (AGP)\cite{colemanAGP}. 
It can be shown that for a singlet type geminal with a fixed number of particles 
the AGP and BCS wave functions are identical\cite{Schrieffer}
and can be expressed in the form of single determinant\cite{bouchaud1}.  


The BCS wave function has been recently used to calculate several atoms and molecules as well as  
superfluid Fermi gases~\cite{sorellabcs1, sorellabcs2,carlsonbcs}.
The results show promising gains when compared to 
the single-determinant Hartree--Fock (HF) wave functions, 
nevertheless, in partially
spin-polarized systems the improvements are
less pronounced due to the lack of pair correlations in
the spin-polarized subspace~\cite{sorellabcs1, sorellabcs2}. 
The spin-polarized (triplet) pairing wave functions naturally lead to pfaffians (instead of 
determinants). In this respect, pfaffian have been mentioned a few times and applied to model
systems~\cite{bouchaud2, Bhattacharjee, kevin} in the past. 

In this paper, we follow upon our previous letter~\cite{pfaffianprl}, in which
we have proposed the description of electron systems by pfaffian wave functions 
with variational freedom beyond HF and BCS wave functions.
The pfaffian form proved to be the right algebraic form,
which can accommodate pair orbitals with 
singlet and triplet pair channels, together with 
unpaired one-particle orbitals, into a single compact wave function.
Here we present a set of key mathematical identities
and formulas for pfaffians, some of them derived for the first time. 
We investigate generalizations to linear combinations 
of pfaffians and to antisymmetrized independent singlet pairs and compare 
the results from the point of view of recovered energies and compactness of
the wave functions. We analyze the fermion nodes for some of the wave functions
and point out the topological differences between HF, pfaffian and 
essentially exact wave functions for a given test example. Finally, 
we explore the possibility of further improvements of nodal structure of pfaffians by 
using the recently proposed form of backflow correlations~\cite{drummond_bf,rios_bf}.

\section{Algebra of pfaffians}
\subsection{Definitions}
First introduced by Arthur Cayley in 1852~\cite{cayley2},  the pfaffian is named after 
German mathematician Johann Friedrich Pfaff.
Given a $2n\times 2n$ skew-symmetric matrix $A=\left[a_{i,j}\right]$, the 
pfaffian of $A$ is defined as antisymmetrized product
\begin{align}\label{eq:pfdef}
{\rm pf}[A]&={\cal A}[a_{1,2}a_{3,4}\ldots a_{2n-1,2n}] \nonumber \\
&=\sum_{\alpha} {\rm sgn}(\alpha)\ a_{i_1, j_1}a_{i_2, j_2} \ldots a_{i_n, j_n},
\end{align}  
where the sum runs over all possible $(2n-1)!!$ pair partitions 
$\alpha=\{(i_1, j_1),(i_2, j_2),\ldots ,(i_n, j_n))\}$ of $\{1,2,\ldots,2n\}$ with $i_k < j_k$.
The sign of permutation associated with the partition $\alpha$ is denoted as
${\rm sgn}(\alpha)$. 
The pfaffian for a matrix of odd order equals to zero. The following example gives pfaffian
of a  $A(4 \times 4)$ skew-symmetric matrix:
\begin{align}
{\rm pf}\begin{bmatrix}
0 & a_{12} & a_{13} & a_{14} \\
-a_{12} & 0 & a_{23} & a_{24} \\
-a_{13} & -a_{23} & 0 &  a_{34} \\
-a_{14} & -a_{24} & -a_{34} & 0\\
\end{bmatrix}=a_{12}a_{34}-a_{13}a_{24}+a_{14}a_{23}.
\end{align}   
It can be also evaluated recursively as 
\begin{align}\label{eq:pfrecurr}
{\rm pf}[A]&=\sum_{j=2}^{2n} a_{1,j}\sum_{\alpha_{1,j}}{\rm sgn}(\alpha_{1,j})\ a_{i_1, j_1}a_{i_2, j_2} \ldots a_{i_{n-1}, j_{n-1}}\nonumber\\ 
&\equiv \sum_{j=2}^{2n} a_{1, j} P_c(a_{1, j}) ,
\end{align} 
where $\alpha_{1,j}$ is partition with $i_k, j_k\neq 1, j$ and $P_c(a_{1, j})$ is 
defined as pfaffian cofactor of $a_{1, j}$.
The cofactor for an element $a_{j,k}$ is given by a formula 
\begin{align}\label{eq:pfcof}
P_c(a_{j,k})=(-1)^{j+k+1}{\rm pf}[A(j,k;j,k)],  
\end{align}
where the matrix $A(j,k;j,k)$ has the rank $2(n-1)\times 2(n-1)$  and is obtained  from  $A$ by eliminating
rows $j$ and $k$ and columns  $j$ and $k$. 
\subsection{Calculation of a pfaffian}
There exist several identities involving pfaffians and determinants.
For any  $2n \times 2n$ skew-symmetric matrix $A$ and arbitrary matrices $B(2n \times 2n)$ 
and $M(n \times n)$ we have the following relations:
\begin{subequations}
\begin{align}
{\rm pf}[A^T]&=(-1)^n {\rm pf}[A] \label{eq:pfident1}\\
{\rm pf}[A]^2&={\rm det}[A] \label{eq:pfident2}\\
{\rm pf} \left[\begin{array}{cc}
A_1 & 0 \\
0 & A_2
\end{array} \right]&={\rm pf}[A_1]{\rm pf}[A_2] \label{eq:pfident5}\\
{\rm pf}[BAB^T]&={\rm det}[B]{\rm pf}[A]\label{eq:pfident4} \\
{\rm pf}\left[\begin{array}{cc}
0 & M \\
-M^T & 0
\end{array} \right]
&=(-1)^{{n(n-1)}\over{2}}{\rm det}[M]\label{eq:pfident3}
\end{align}
\end{subequations}
Proofs:
 
(\ref{eq:pfident1}) Each permutation contains product of 
$n$ pairs resulting in an overall $(-1)^n$ factor.

(\ref{eq:pfident2}) This is a well-known Cayley's relationship between
the pfaffian and the determinant of a skew-symmetric
matrix. Since it has been proved many times before in variety of ways~\cite{nakahara,sss,cayley},  
we do not give this proof here. Using this relation we rather prove
 a more general version of Cayley's identity~\cite{cayley}
in the Appendix ~(\ref{appendix:Cayley}), which we were not able to find 
anywhere else except in the original Cayley's paper~\cite{cayley}.

(\ref{eq:pfident5}) Use the expansion by pfaffian cofactors.

(\ref{eq:pfident4}) By squaring (4d), using Eq.~(\ref{eq:pfident2}), and taking the
square root one finds
${\rm pf}[BAB^T]=\pm{\rm det}[B]{\rm pf}[A]$. Substituting the identity matrix $I$ for $B$ 
one finds $+$ to be the correct sign.

(\ref{eq:pfident3}) Assume
\begin{align*}
B=\left(\begin{array}{cc}
M & 0\\
0 & I
\end{array}\right)
\quad {\rm and} \quad
A=\left(\begin{array}{cc}
0 & I\\
-I & 0
\end{array}\right)
\end{align*}
in Eq.~(\ref{eq:pfident4}). The overall sign is given by value of ${\rm pf}[A]$.

The identities listed above imply several important properties.
First, Eqs.~(\ref{eq:pfident4}) and (\ref{eq:pfident3}) show that every determinant 
can be written as a pfaffian, but on the contrary, 
only the absolute value of pfaffian can be given by determinant [Eq.~(\ref{eq:pfident2})].
The pfaffian is therefore a generalized form of the determinant.
Second, by substituting suitable matrices\cite{matrices} for $M$
in Eq.~(\ref{eq:pfident4}) one can verify the following three  
properties of pfaffians \cite{galbiati}, similar to the well-known
properties of determinant:
\begin{itemize}
\item[(a)] multiplication of a row and a column by a constant
is equivalent to multiplication of pfaffian by the same constant,  
\item[(b)] simultaneous interchange of two different rows and corresponding columns changes the sign of pfaffian,
\item[(c)] a multiple of a row and corresponding column added to 
to another row and corresponding column does not change the value of pfaffian.
\end{itemize}
Any real skew-symmetric matrix can be brought to block-diagonal form by an orthogonal transformation.
Recursive evaluation [Eq.~(\ref{eq:pfrecurr})] then implies that the pfaffian of block-diagonal matrix is directly given by 
\begin{align}
\small
{\rm pf}\begin{bmatrix}
0 & \lambda_1  & & & &  \\
-\lambda_1 & 0 & & & & 0 & \\
& & 0 & \lambda_2 & & \\
& & -\lambda_2 & 0 & & \\
& & & &  \ddots & \\
& 0 & & & & 0 & \lambda_n \\
& & & & & -\lambda_n & 0  \\
\end{bmatrix}=\lambda_1 \lambda_2 \ldots \lambda_n.
\end{align}
Therefore by employing a simple Gaussian elimination technique with row pivoting we can 
transform any skew-symmetric matrix into block-diagonal form and obtain 
its pfaffian value in $O(n^3)$ time.

However, in QMC applications one often needs to 
evaluate the wave function after a single electron
update. Since Cayley~\cite{cayley} showed [for proof see App.~(\ref{appendix:Cayley})]
that  
\vspace{0.5cm}
\begin{widetext}
\begin{align}\label{eq:cayley}
\small
{\rm det} \left[\begin{array}{ccccc}
0  & b_{12}  & b_{13} &\ldots &  b_{1,n}\\
-a_{12}  & 0  & a_{23} & \ldots &  a_{2,n}\\
-a_{13}  & -a_{23} & 0  & \ldots &  a_{3,n}\\
 \vdots & \vdots & \vdots &  \ddots &  \vdots \\
-a_{1,n} & -a_{2,n} & -a_{3,n} & \ldots &  0\\
\end{array}\right]=
{\rm pf} \left[\begin{array}{cccccc}
0  & a_{12}  & a_{13} &\ldots &  a_{1,n}\\
-a_{12}  & 0  & a_{23} & \ldots &  a_{2,n}\\
-a_{13}  & -a_{23} & 0  & \ldots &  a_{3,n}\\
 \vdots & \vdots & \vdots &  \ddots &  \vdots \\
-a_{1,n} & -a_{2,n} & -a_{3,n} & \ldots &  0\\
\end{array}\right]\,
{\rm pf} \left[\begin{array}{ccccc}
0  & b_{12}  & b_{13} &\ldots &  b_{1,n}\\
-b_{12}  & 0  & a_{23} & \ldots &  a_{2,n}\\
-b_{13}  & -a_{23} & 0  & \ldots &  a_{3,n}\\
 \vdots & \vdots & \vdots &  \ddots &  \vdots \\
-b_{1,n} & -a_{2,n} & -a_{3,n} & \ldots &  0\\
\end{array}\right],
\end{align}
\end{widetext}
we can relate the pfaffian of original matrix ${\rm pf}[A]$ to  
the pfaffian of a matrix with updated first row and column ${\rm pf}[B]$ using the inverse 
matrix $A^{-1}$ in only $O(n)$ operations by 
\begin{align}\label{eg:inverseupdate}
{\rm pf}[B]=\frac{{\rm det}[A]\sum_j b_{1j}A^{-1}_{j1}}{{\rm pf}[A]}={\rm pf}[A]\sum_j b_{1j}A^{-1}_{j1}.
\end{align}
The second part of Eq.~(\ref{eg:inverseupdate}) was obtained by 
taking advantage of the identity in Eq.~(\ref{eq:pfident2}).
Slightly more complicated relation between ${\rm pf}[A]$ and ${\rm pf}[B]$ 
can be derived if one considers simultaneous change of two separate rows/columns,
which represents a two electron update of the wave function.

\subsection{Gradient and Hessian of pfaffian}
If the elements of matrix $A$ depend on some variational parameters $\{c_i\}$, 
one can derive the following useful expressions (see Sec.~\ref{sec:4}) for gradient and Hessian of pfaffian:
\begin{align}\label{eq:gradient}
\frac{1}{{\rm pf}[A]}\frac{\partial {\rm pf}[A]}{\partial c_i}=\frac{1}{2} {\rm tr}
\left[ A^{-1}\frac{\partial A}{\partial c_i}\right]
\end{align}
and
\begin{align}\label{eq:hessian}
\frac{1}{{\rm pf}[A]}\frac{\partial^2{\rm pf}[A]}{\partial c_i \,\partial c_j}=
&\frac{1}{2} {\rm tr} \left[  A^{-1}\frac{\partial^2 A}{\partial c_i \partial c_j} \right]-\frac{1}{2} {\rm tr} \left[  A^{-1}\frac{\partial A}{\partial c_i}A^{-1}\frac{\partial A}{\partial c_j} \right] \nonumber \\
&+\frac{1}{4}{\rm tr}\left[A^{-1}\frac{\partial A}{\partial c_i}\right] \, {\rm tr}\left[ A^{-1}\frac{\partial A}{\partial c_j}\right] ,
\end{align}
where 
$A^{-1}$ is again the inverse of $A$.

\section{\label{sec:level2} Pairing wave functions}
In order to contrast the properties of pair wave functions with the wave functions build from
one-particle orbitals we will first recall 
the well-known fact from the Hartree-Fock theory. The simplest antisymmetric
wave function for $N$ electrons constructed from one-particle orbitals is the Slater {\em determinant}
\begin{align}\label{eq:hfslater}
\Psi_{HF}= {\rm det} [\tilde\varphi_k({\bf r_i},s_i)]={\rm det} [\tilde\varphi_k(i)]; \quad i,k=1,\ldots,N,
\end{align}
where tilde means that the one-particle states depend on
both space and spin variables. Clearly, for $N$ electrons,
this requires $N$ linearly independent spin-orbitals which form an orthogonal
set in canonical HF formulation. 

Let us now consider the generalization of the one-particle orbital
to a two-particle (or pair) orbital
$\tilde{\phi}(i,j)$, where tilde again denotes dependence on both
spatial and spin variables. The simplest antisymmetric wave function 
for $2N$ electrons constructed from the pair orbital is a
{\em pfaffian} 
\begin{align}\label{eq:generalpairingwf}
\Psi={\cal A}[\tilde{\phi}(1,2),\tilde{\phi}(3,4) \ldots\tilde{\phi}(2N-1,2N)]={\rm pf}[\tilde{\phi}(i,j)].
\end{align}
The antisymmetry is guaranteed by the definition (\ref{eq:pfdef}), since the signs 
of pair partitions alternate depending on the parity of the corresponding
permutation. 
The important difference from Slater determinant is that in the simplest case only {\em one} 
pair orbital is necessary. (This can be generalized, of course, as will be shown later.)
If we further restrict our description to systems with collinear spins, 
the pair orbital $\tilde{\phi}({\bf r}_i, s_i; {\bf r}_j, s_j)$
for two electrons in positions ${\bf r}_i$ and ${\bf r}_j$ 
and with spins projections $s_i$ and $s_j$ and can be expressed as 
\begin{align}\label{eq:generalpair}
\tilde{\phi}({\bf r}_i, s_i; {\bf r}_j, s_j)&=
\phi(i,j)
\langle s_is_j|[|\uparrow \downarrow\rangle -|\downarrow\uparrow\rangle]/\sqrt{2}\\ \nonumber
&+\chi^{\uparrow \uparrow}(i,j)\langle s_is_j|\uparrow\uparrow\rangle\\ \nonumber
&+\chi^{\uparrow \downarrow}(i,j)
\langle s_is_j|[|\uparrow \downarrow\rangle +|\downarrow\uparrow\rangle]/\sqrt{2}\\ \nonumber
&+\chi^{\downarrow \downarrow}(i,j)\langle s_is_j|\downarrow\downarrow\rangle.
\end{align}
Here  
$\phi(i,j)=\phi({\bf r}_i,{\bf r}_j)$ is even, while $\chi^{\uparrow \uparrow}$, $\chi^{\uparrow \downarrow}$ and $\chi^{\downarrow \downarrow}$ are 
odd functions of spatial coordinates. 
In the rest of this section we will discuss special cases of wave function (\ref{eq:generalpairingwf}).

\subsection{Singlet pairing wave function}\label{subsec:singlet}
Let us consider the first $i=1,2, ...,N$ electrons to be spin-up and the rest $j=N+1, ...,2N$ 
electrons to be spin-down and allow only $\phi({\bf r}_i,{\bf r}_j)$ in $\tilde{\phi}({\bf r}_i, s_i; {\bf r}_j, s_j)$ to be non-zero. 
Using the pfaffian identity [Eq.~(\ref{eq:pfident3})] we can write 
the wave function for $N$ singlet pairs, also known as the BCS wave function, in the following form
\begin{align}
\Psi_{BCS}={\rm pf}\begin{bmatrix}
0 & 
{\boldsymbol \Phi}^{\uparrow\downarrow}\\
-{\boldsymbol \Phi}^{\uparrow\downarrow T}&
0 \\
\end{bmatrix}={\rm det}[\boldsymbol \Phi^{\uparrow\downarrow}],
\end{align}
which is simply a determinant of the $N\times N$ matrix $\boldsymbol \Phi^{\uparrow\downarrow}=\left[\phi (i,j)\right]$ as was shown
previously\cite{bouchaud1}.

It is straightforward to show that the BCS wave function contains the 
restricted HF wave function as a special case.
Let us define the Slater matrix ${C}=\left[\varphi_i(j)\right]$, where $\{\varphi_i\}$ is a set of HF occupied
orbitals. Then we can write 
\begin{align}
\Psi_{HF}={\rm det}[C]{\rm det}[C]=
    {\rm det}[CC^T]={\rm det}[{\boldsymbol \Phi}_{HF}^{\uparrow\downarrow}],
\end{align}
where 
\begin{align}
( \boldsymbol \Phi_{HF}^{\uparrow\downarrow})_{i,j}=\phi_{HF}(i,j)=\sum_{k=1}^{N}\varphi_k(i)\varphi_k(j).
\end{align}
On the other hand, we can think of the BCS wave function as a natural generalization
of the HF one. To do so we write the singlet pair orbital as
\begin{align}\label{eq:phi}
\phi(i,j)=\sum_{k,l}^{>N}S_{k,l}\varphi_k(i)\varphi_l(j)=
{\boldsymbol \varphi}(i)\,{\bf S}\,{\boldsymbol \varphi}(j),
\end{align}
where the sum runs over all (occupied and virtual) 
single-particle orbitals and ${\bf S}$ is some symmetric matrix. 
Therefore, we can define one-particle orbitals which diagonalize this matrix and call them 
\emph{natural orbitals of a singlet pair}.


The BCS wave function is efficient for describing
systems with single-band correlations such
as Cooper pairs in conventional BCS
superconductors where pairs form from
one-particle states close to the Fermi level.

\subsection{Triplet pairing wave function}\label{subsec:triplet}
Let us assume, in our system of $2N$ electrons, that the first $M_1$ are spin-up 
and remaining $M_2=2N-M_1$ are spin-down. Further we restrict $M_1$ and $M_2$ to be even numbers.  
Then by allowing only $\chi^{\uparrow \uparrow}(i,j)$ and $\chi^{\downarrow \downarrow}(i,j)$
in (\ref{eq:generalpair}) to be non-zero, we obtain from (\ref{eq:generalpairingwf})
by the use of Eq.~(\ref{eq:pfident5})
\begin{align}
\Psi_{T}={\rm pf}\begin{bmatrix}
 {\boldsymbol \xi}^{\uparrow\uparrow} & 
0\\
0 & {\boldsymbol \xi}^{\downarrow\downarrow}\\
\end{bmatrix}={\rm pf}[{\boldsymbol \xi}^{\uparrow\uparrow}]{\rm pf}[{\boldsymbol \xi}^{\downarrow\downarrow}],
\end{align}
where we have introduced $M_1\times M_1(M_2\times M_2)$ matrices 
${\boldsymbol \xi}^{\uparrow\uparrow(\downarrow\downarrow)}=\left[\chi^{\uparrow\uparrow(\downarrow\downarrow)}(i,j)\right]$.
To our knowledge, this result was never explicitly stated and
only the weaker statement that the square of wave function simplifies to a product of determinants 
has been given.\cite{{bouchaud1}}

The connection to a restricted HF wave function for the above state can be again 
established as follows.
In accord with what we defined above,
 ${\rm det}[C^{\uparrow(\downarrow)}]$ are
spin-up(-down) Slater determinants of some HF orbitals $\{\varphi_i\}$.
Then, by taking advantage of  Eq.~(\ref{eq:pfident3}) we can write
\begin{align}
\Psi_{HF}&={\rm det}[C^{\uparrow}]{\rm det}[C^{\downarrow}]\\ \nonumber
&=\frac{{\rm pf}[C^{\uparrow} A_1 {C^{\uparrow}}^T]{\rm pf}[C^{\downarrow} A_2 {C^{\downarrow}}^T]}{{\rm pf}[A_1]{\rm pf}[A_2]},
\end{align}
given $A_1$ and $A_2$ are some skew-symmetric non-singular matrices.
In the simplest case, when $A_1$ and $A_2$ have block-diagonal form with all values $\lambda_i=1$,
one gets
\begin{align}
\Psi_{HF}={\rm pf}[\boldsymbol \xi_{HF}^{\uparrow\uparrow}]{\rm pf}[\boldsymbol \xi_{HF}^{\downarrow\downarrow}].
\end{align}
The pair orbitals can be then expressed as
\begin{align}
(\boldsymbol \xi_{HF}^{\uparrow\uparrow(\downarrow\downarrow)})_{i,j}&=\chi_{HF}^{\uparrow\uparrow(\downarrow\downarrow)}(i,j)\\ \nonumber
&=\sum_{k=1}^{M_1(M_2)/2} [\varphi_{2k-1}(i)\varphi_{2k}(j)-\varphi_{2k-1}(j)\varphi_{2k}(i)].
\end{align} 
Similarly to the singlet pairing case, one can also think of triplet pairing 
as a natural generalization of the HF wave function. To do so we write the triplet pair orbitals as
\begin{align}\label{eq:chi}
\chi(i,j)^{\uparrow\uparrow(\downarrow\downarrow)}&=\sum_{k,l}^{>M_1(M_2)}A^{\uparrow\uparrow(\downarrow\downarrow)}_{k,l}\varphi_k(i)\varphi_l(j) \nonumber \\
&={\boldsymbol \varphi}(i)\,{\bf A}^{\uparrow\uparrow(\downarrow\downarrow)}\,{\boldsymbol \varphi}(j),
\end{align}
where again the sum runs over all (occupied and virtual) single-particle orbitals and ${\bf A}^{\uparrow\uparrow(\downarrow\downarrow)}$ are some skew-symmetric
matrices. Therefore, we can define one-particle orbitals which block-diagonalize these matrices and call them 
\emph{natural orbitals of a triplet spin-up-up(down-down) pair}. 

\subsection{Generalized pairing wave function}
Let us now consider a partially spin-polarized system with
unpaired electrons. In order to introduce both types 
of pairing we allow $\chi^{\uparrow \uparrow}(i,j)$, $\chi^{\downarrow \downarrow}(i,j)$
and $\phi(i,j)$ in (\ref{eq:generalpair}) to be non-zero. 
However, we omit the $\chi^{\uparrow \downarrow}(i,j)$ term. 
Then our usual ordered choice of electron labels, with all spin-up electrons first 
and remaining electrons spin-down, enables us to directly write from (\ref{eq:generalpairingwf}) the
singlet--triplet--unpaired (STU) orbital pfaffian wave function~\cite{pfaffianprl}
\begin{align}
\Psi_{STU}=
{\rm pf}\begin{bmatrix}
{\boldsymbol \xi}^{\uparrow\uparrow} & 
{\boldsymbol \Phi}^{\uparrow\downarrow} & 
{\boldsymbol\varphi}^{\uparrow} \\
-{\boldsymbol \Phi}^{\uparrow\downarrow T} &
{\boldsymbol \xi}^{\downarrow\downarrow} &
{\boldsymbol \varphi}^{\downarrow} \\
-{\boldsymbol\varphi}^{\uparrow T} &
-{\boldsymbol\varphi}^{\downarrow T} &
0 \;\; \\
\end{bmatrix},
\end{align} 
where the bold symbols are block matrices and vectors of
corresponding orbitals as defined in sections~\ref{subsec:singlet}, \ref{subsec:triplet} 
and $T$ denotes transposition. Let us note that for a spin-restricted 
STU wave function the pair and one-particle orbitals of spin-up and -down channels
would be identical. 

The above pfaffian form can accommodate both singlet and triplet pairs as well as one-particle
unpaired orbitals into a single, compact wave function.
The correspondence of STU pfaffian wave function to HF wave function can 
be established in a similar way to the pure singlet and triplet pairing cases.

\section{Pairing Wave function Results}\label{sec:4}
We perform the variational and fixed-node diffusion Monte Carlo
(VMC and DMC) calculations \cite{hammond,qmcrev} with the pfaffian pairing wave functions.
As we mentioned earlier, the trial variational wave function is a product
of an antisymmetric part $\Psi_A$
times the Jastrow correlation factor
\begin{align}\label{eq:slater-jastrowwf}
\Psi_T ({\bf R}) = \Psi_{A}({\bf R}) \exp[U_{corr}(\{r_{ij}\},\{r_{iI}\},
\{r_{iJ}\})],
\end{align}
where  $U_{corr}$ depends on
electron-electron, electron-ion and, possibly, on 
electron-electron-ion combinations of distances \cite{qmcrev,cyrus2,qwalk}.
Previously, we have reported\cite{pfaffianprl} the results with antisymmetric part
being equal to $\Psi_A=\Psi_{HF}$, $\Psi_A=\Psi_{BCS}$ and 
$\Psi_A=\Psi_{STU}$. We extend this work to different linear combinations of pfaffians.
The pair orbitals were expanded in products of a one-particle orbital
basis according to Eqs.~(\ref{eq:phi}, \ref{eq:chi}).
The expansions include both occupied and  virtual one-particle orbitals
from either Hartree-Fock or CI correlated calculations\cite{szabo}.
The pair orbital expansion coefficients were then optimized
in VMC by minimizations of energy, variance or a combination
of energy and variance~\cite{cyrus2}. The optimization procedure 
requires the calculation of gradient and the Hessian of the wave function 
according to Eqs.~(\ref{eq:gradient}, \ref{eq:hessian}). We used pseudopotentials\cite{lester} 
to eliminate the atomic cores. 

\subsection{Single- and multi-pfaffian calculations}
\begingroup
\begin{table*}[!th]
\caption{Total energies for C and N dimers with 
amounts of correlation energy recovered in VMC and DMC methods
with wave functions as discussed in the text.
The corresponding number of pfaffians/determinants $n$ for each
wave function is also shown. 
See caption of Table~\ref{energies:1} for a more detailed description.}
\begin{ruledtabular}
\begin{tabular}{l l c c c c c c}
\multicolumn{1}{l}{Method}& \multicolumn{1}{l}{WF} & \multicolumn{1}{c}{$n$} & \multicolumn{1}{c}{C$_2$} &  \multicolumn{1}{c}{E$_{corr}$[\%]} &  \multicolumn{1}{c}{$n$} &  
\multicolumn{1}{c}{N$_2$} & \multicolumn{1}{c}{E$_{corr}$[\%]}\\
\hline
VMC & MPF                 &  5   &  -11.0187(2) & 87.8(1) &  5   &  -19.8357(3)   &  89.2(1)\\
    & APG                 &  4!  &  -11.0205(4) & 88.3(1)  &  5!  &  -19.8350(3)   &  89.0(1)\\
    & CI\footnotemark[1]  & 148  &  -11.0427(1) & 93.7(1) & 143  &  -19.8463(9)   &  91.6(2)\\
\hline
DMC & MPF                 &  5   &  -11.0437(4) & 94.0(1)  &  5   &  -19.8623(5)   &  95.3(1)\\
    & APG                 &  4!  &  -11.0435(7) & 94.0(2)  &  5!  &  -19.8611(3)   &  95.0(1)\\
    & CI\footnotemark[1]  & 148  &  -11.0573(2) & 97.3(1) & 143  &  -19.875(2)    &  98.3(5)\\
\end{tabular}
\end{ruledtabular}
\footnotetext[1]{The determinantal weights were re-optimized in the VMC method.}
\label{energies3}
\end{table*}
\endgroup
The results of a single-pfaffian wave function calculations 
when applied to the first row atoms and dimers were reported in 
our previous paper\cite{pfaffianprl}. However, for completeness, 
we summarize all total and correlation energies in Table~\ref{energies:1} Appendix~\ref{appendix:table}.
A systematic high percentage of recovered correlation energy on the level of 94-97\% in DMC method 
with generally low triplet contributions was observed.

Further, the tests of multi-pfaffian (MPF) wave function of the form
\begin{align}\label{eq:mpf}
\Psi_{MPF}= 
{\rm pf}[\chi^{\uparrow\uparrow}_1,\chi^{\downarrow\downarrow}_1,\phi_1,\varphi_1]+
{\rm pf}[\chi^{\uparrow\uparrow}_2,\chi^{\downarrow\downarrow}_2,\phi_2,\varphi_2]+\ldots
\end{align}
for atomic systems were also discussed in our previous study\cite{pfaffianprl}. 
A small number of pfaffians was shown to recover another significant fraction of the missing 
correlation energy comparable to much more extensive configuration interaction expansions in determinants.

In this work, we extend the application of MPF wave functions to the diatomic cases (Table~\ref{energies3}).
However, only very limited gain over single STU pfaffian WF was achieved 
for MPF wave functions with few pfaffians. 
We therefore conclude that for obtaining significantly larger gains in correlation energy
the molecular wave functions require much larger expansions. 

\subsection{APG wave function}
We have also tested the fully antisymmetrized product of geminals (or singlet independent pairs) wave function, 
which introduces one pair orbital per each electron pair.
For system of $2N$ fermions in singlet state, the APG wave function can be written as
\begin{align}
\Psi_{APG}&={\cal A} [\tilde{\phi}_1 (1,2), \tilde{\phi}_2(3,4), \ldots,\tilde{\phi}_N (2N-1,2N)]\\ \nonumber
&=\sum_{P}{\rm pf}[\tilde{\phi}_{i_1}(1,2), \tilde{\phi}_{i_2}(3,4), \ldots,\tilde{\phi}_{i_N} (2N-1,2N)], 
\end{align}
where the last equation corresponds to the sum over all $N!$ possible permutations 
of $N$ different pair orbitals $\tilde{\phi}_{i}$ for each pfaffian. 
Recently, the APG wave function was used also by Rassolov\cite{rassolov} 
in the form of an strongly orthogonal geminals. 
Our results for $C$ and $N$ dimers using APG wave functions in the VMC and DMC methods are given in Table~\ref{energies3}. 

Consideration of independent pairs results in an exponential increase of number of pfaffians.
However, captured correlation energy is on the level of small MPF expansion and 
significantly less than CI with re-optimized weights using the same one-particle orbitals.  
This suggests that to achieve more correlation energy in larger systems, 
we have to go beyond double pairing. 

\subsection{Nodal properties}
\begin{figure*}
   \mbox{
       {\resizebox{2.3in}{!}{\includegraphics{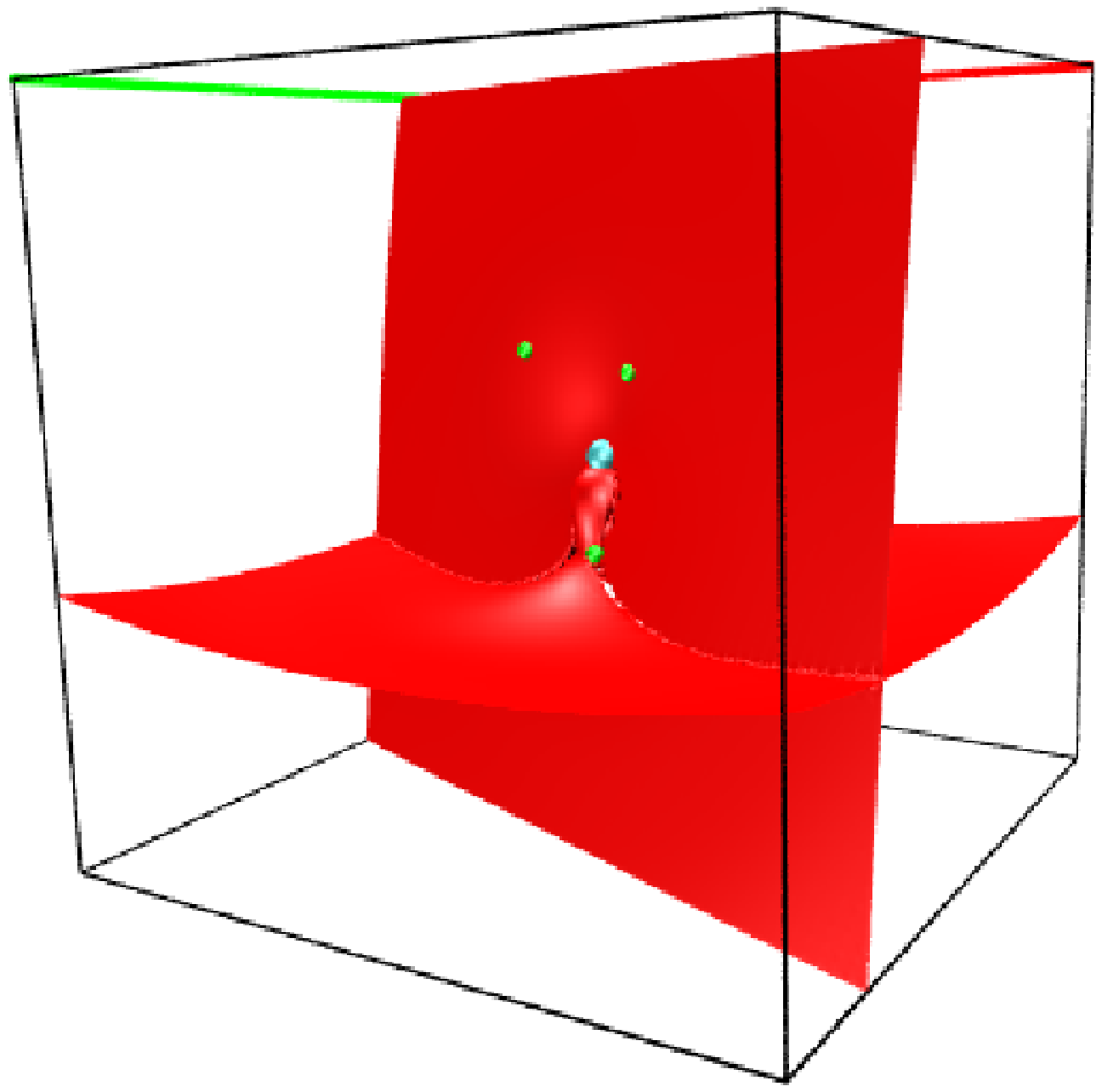}}}\quad
       {\resizebox{2.3in}{!}{\includegraphics{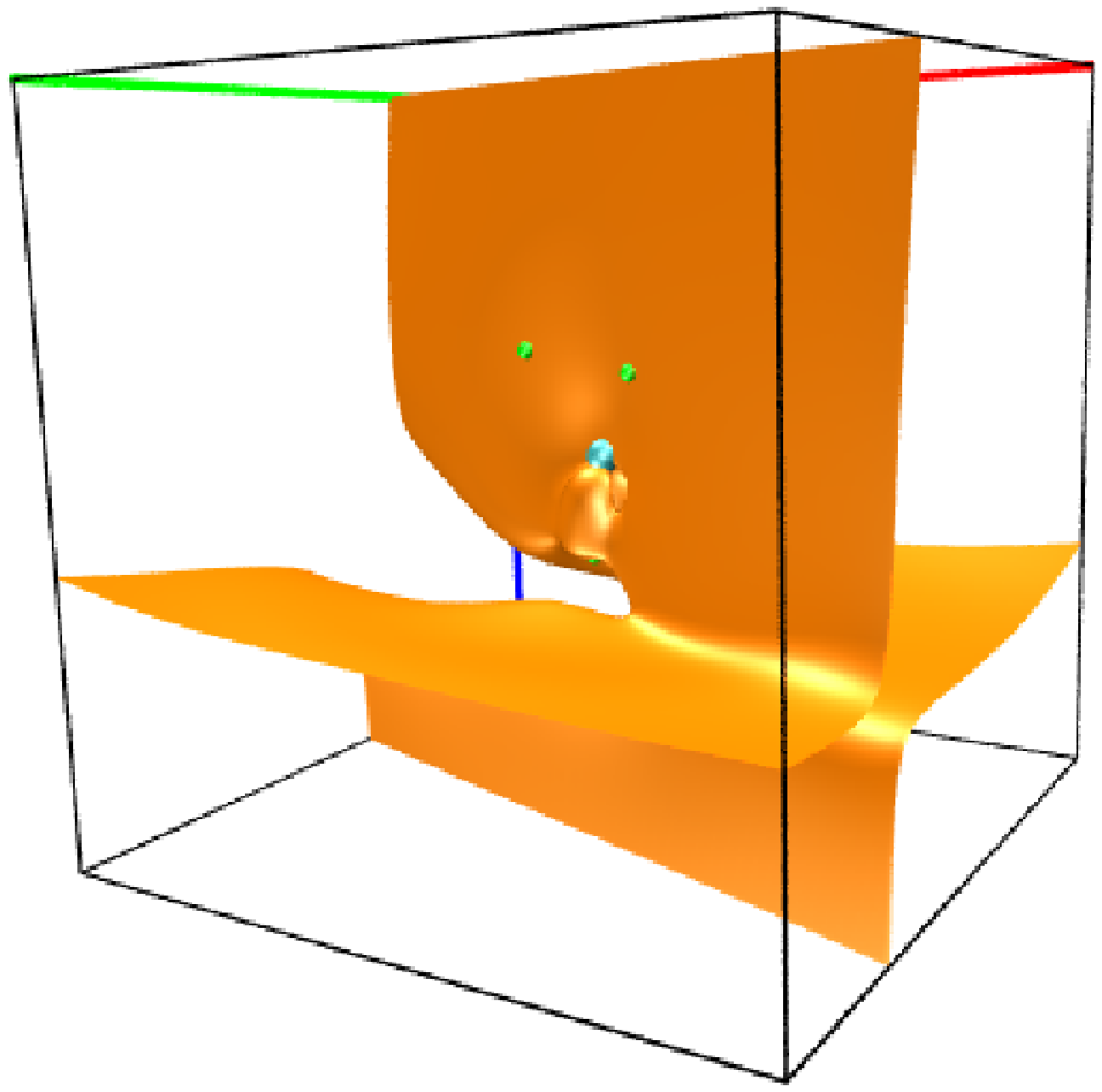}}} \quad
       {\resizebox{2.3in}{!}{\includegraphics{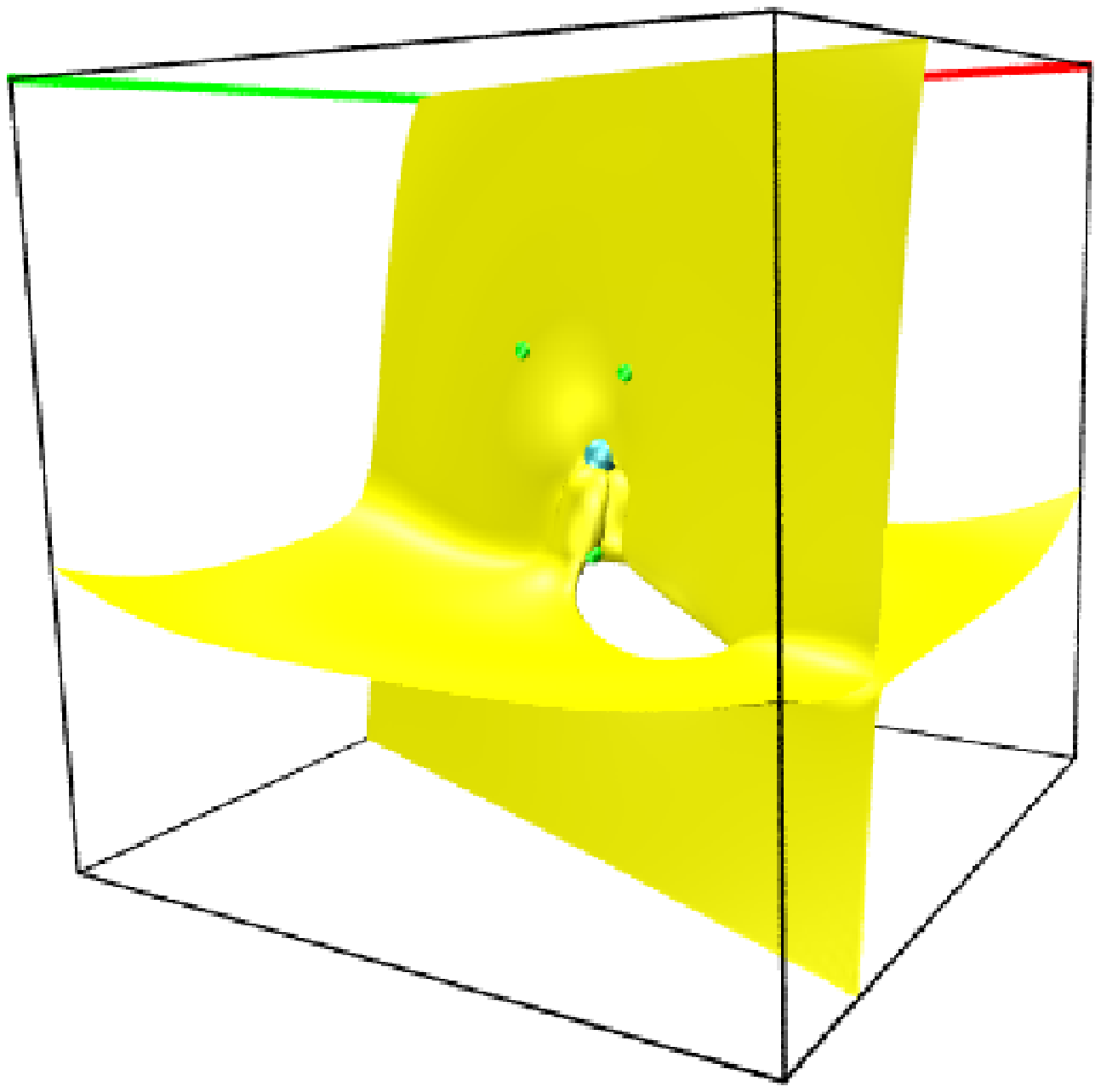}}}
      }
    \mbox{
       {\resizebox{2.3in}{!}{\includegraphics{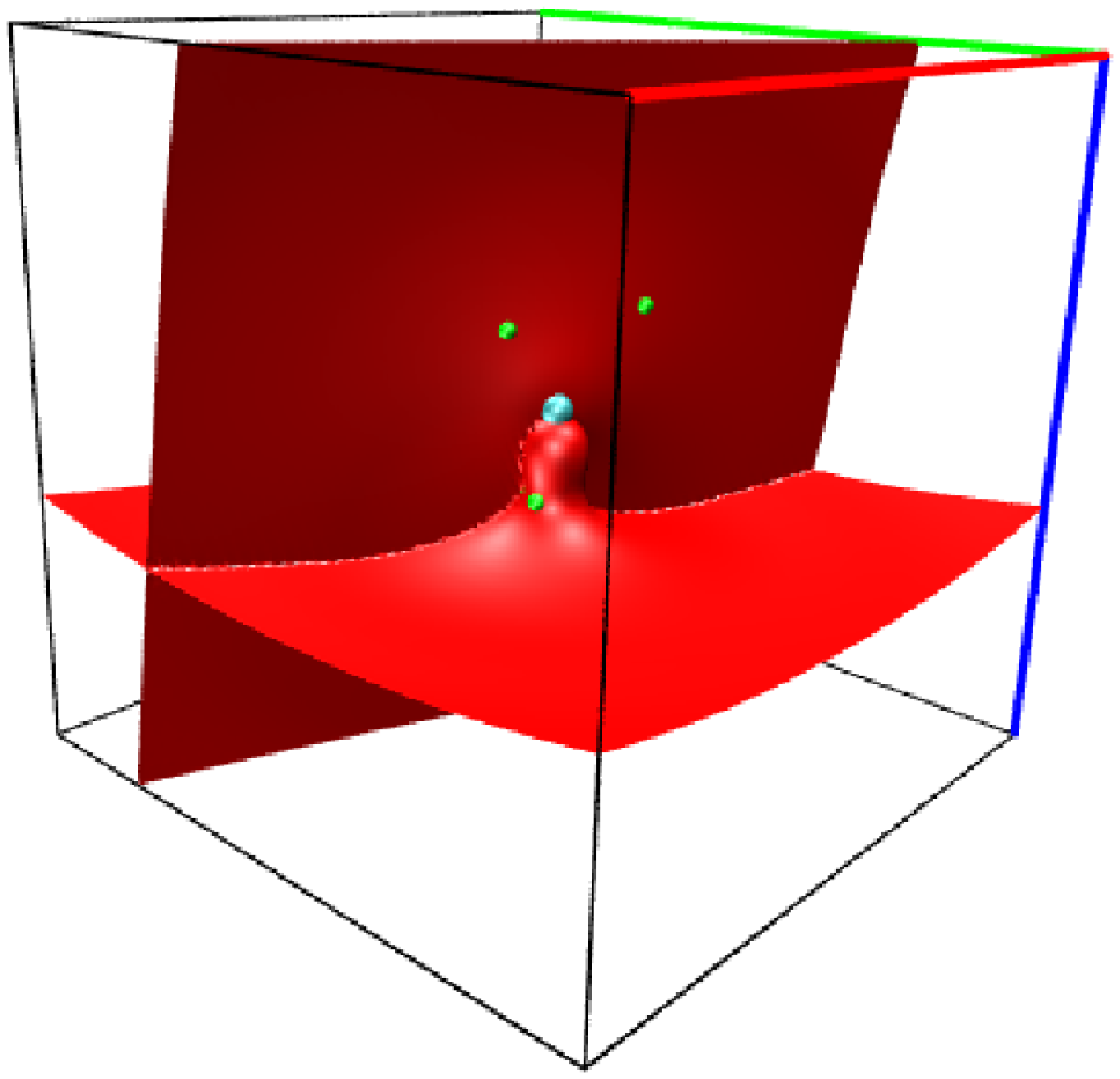}}}\quad
       {\resizebox{2.3in}{!}{\includegraphics{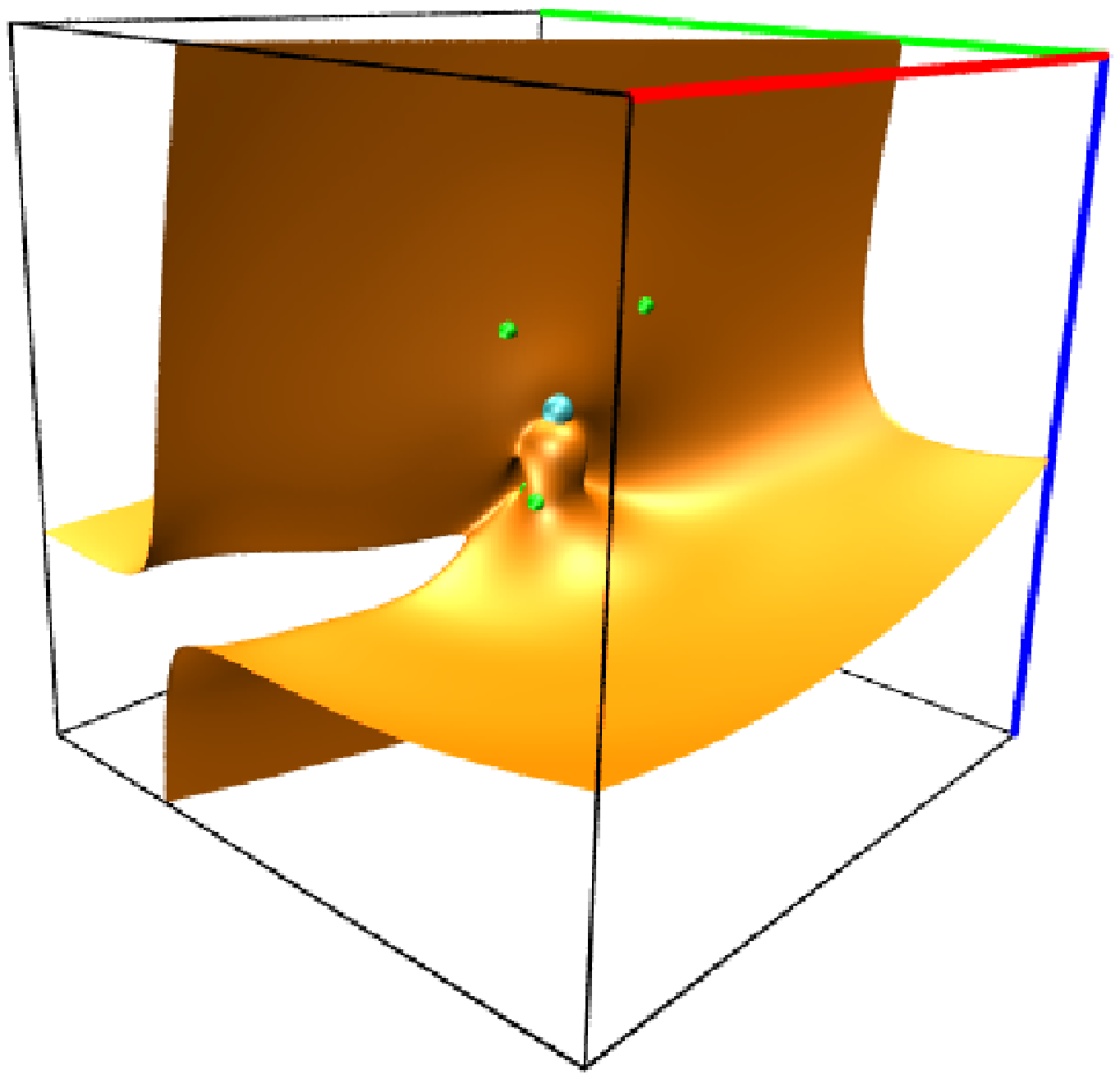}}}\quad
       {\resizebox{2.3in}{!}{\includegraphics{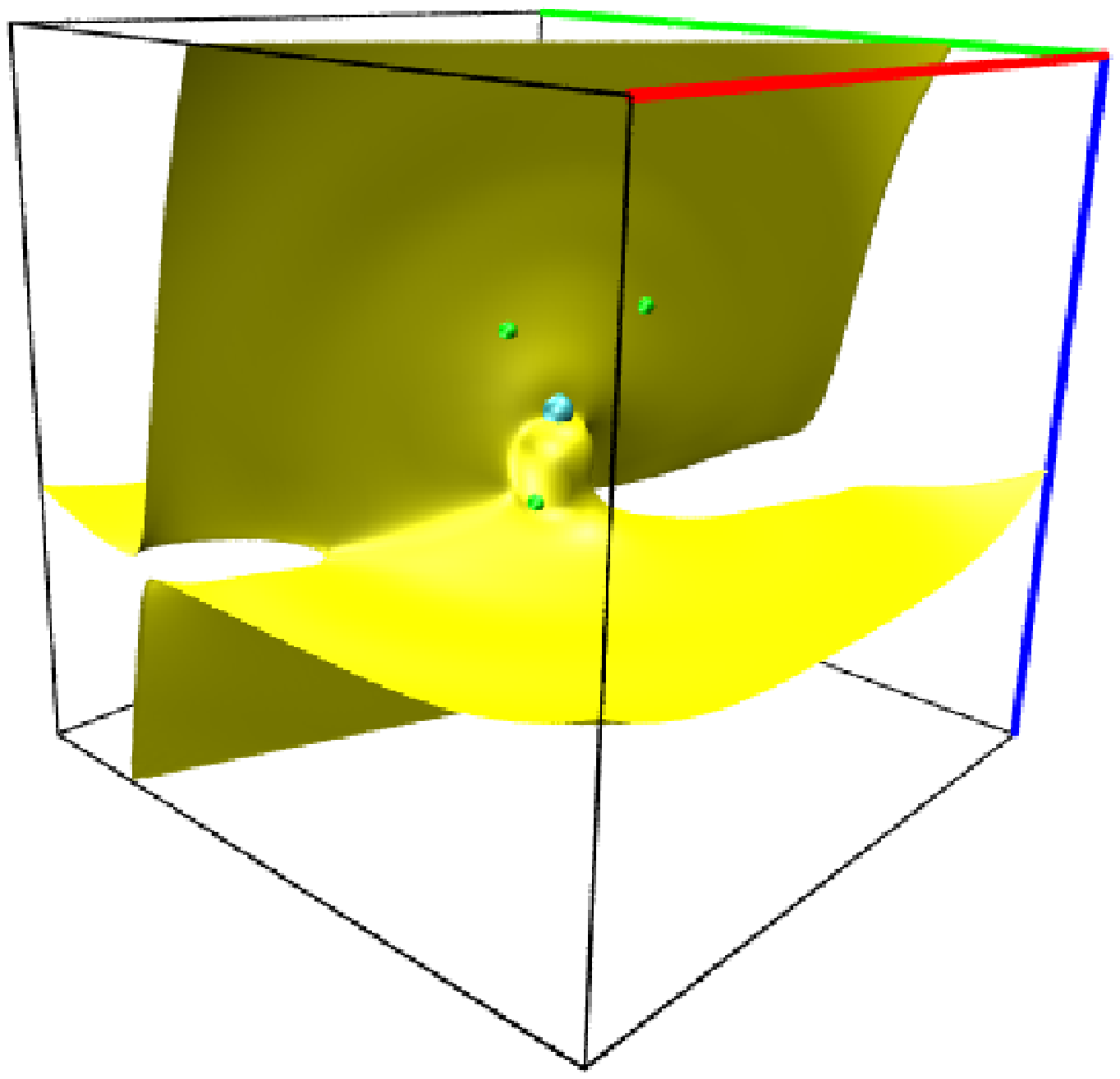}}}
     }
\caption{
(color online) A three-dimensional cut through the fermion node hypersurface of oxygen atom 
obtained by scanning the wave function with a spin-up and -down (singlet)
pair of electrons at equal positions, while keeping the 
rest of electrons at a given VMC snapshot positions (small green spheres). 
Nucleus is depicted in the center of the cube by the blue 
sphere. The three colors (from left to right) show nodes of: Hartree-Fock (red/dark gray);
multi-pfaffian nodes (orange/medium gray); and the nodes of the CI
wave function (yellow/light gray) in two different views (upper and lower rows).
The CI nodal surface is very close to the exact one (see text).
The HF node clearly divides the space into four
nodal cells while pfaffian and CI wave functions
partitioning leads to the minimal number of two nodal cells.
The changes in the nodal topology occur on the appreciable spatial 
scale of the order of 1 a.u.}
\label{fig:nodes2}
\end{figure*}

The fermion node is defined by an implicit equation
$\Psi(R)=0$ and for $N$ electrons it is a \mbox{$(3N-1)$-dimensional} hypersurface.
With exception of few exact cases, the nodes of trial/variational wave functions introduce bias into fixed-node DMC energies. 
Recently a number of authors have reported improvement in nodal structure of trial wave functions 
\cite{Bressanininew, umrigarC2, sorellabcs1, sorellabcs2, pfaffianprl, drummond_bf}.

The effect of pairing correlations on nodes can be highlighted by direct comparison.
Fig. \ref{fig:nodes2} shows the example of nodal structure of oxygen atom. 
Here we compare the nodal surfaces of HF (no pairing), MPF pfaffian (STU pairing) and a
high accuracy CI wave function with more than 3000 determinants, which gives 
essentially exact fermion nodes [i.e., $99.8(3)\%$ of correlation energy in fixed-node DMC].

It is clear that the changes in the nodal surfaces are significant, 
the most important one being the elimination of artificial 
four nodal cells resulting from the independence of spin-up and -down
channels in HF. The pfaffian smooths-out the crossings and fuses
the compartments of the same sign into the single ones. These topology changes
therefore lead to the minimal number of two nodal cells, an effect observed
in correlated context previously \cite{davidnode,dariobe,lubos_nodeprl,lubos_nodeprb}.
However, the nodes of the pfaffian wave functions could be further improved if
the scheme for direct optimization of nodes of trial 
wave functions were used\cite{carlsonbcs,chang2005}. 
Additional result from our work is that despite such a substantial change
in the nodal structure the amount of missing correlation energy is still non-negligible.

\section{Backflow Correlated Wave Functions}
Another route to improvement of the trial wave function and its nodal structure 
is through the introduction of backflow correlations~\cite{feynman,schmidt_bf,panoff,moskowitz,kwon1,kwon2,kwon3,kwon4,markus,drummond_bf,rios_bf}. 
Given the form of our trial wave function, Eq.~(\ref{eq:slater-jastrowwf}), its nodal structure is completely defined 
by the nodes of the antisymmetric part $\Psi_A({\bf R})$. The backflow correlations are then introduced by replacing
$\Psi_A({\bf R})$ by $\Psi_A({\bf X})$, where ${\bf X}=({\bf x}_1,{\bf x}_2,\ldots)$ are some quasi-coordinates 
dependent on all electron positions ${\bf R}$, such that overall antisymmetry is preserved. 
Consequently, if ${\bf X}$ is made dependent on some variational parameters, we can decrease the fixed-node errors by further optimizing $\Psi_A({\bf X})$.

The implementation of the backflow correlations into Slater determinant and pfaffian wave functions 
closely follows the approach of Kwon {\it et al.\/}\cite{kwon2} and Rios {\it et al\/}\cite{rios_bf}. The quasi-coordinate of $i$-th electron at position ${\bf r}_i$ is given as 
\begin{align}\label{eg:bfdisplacement}
{\bf x}_i&={\bf r}_i+{\boldsymbol \xi}_i({\bf R}) \nonumber \\
&={\bf r}_i+{\boldsymbol \xi}_i^{en}({\bf R})+{\boldsymbol \xi}_i^{ee}({\bf R})+{\boldsymbol \xi}_i^{een}({\bf R}),
\end{align}
where ${\boldsymbol \xi}_i$ is the $i$-th electron's backflow displacement 
divided to the contributions from one-body (electron-nucleus), two-body (electron-electron) 
and three-body (electron-electron-nucleus) terms. 
They can be further expressed as 
\begin{align}\label{eg:bfterms}
{\boldsymbol \xi}_i^{en}({\bf R})&=\sum_I \mu(r_{iI}) {\bf r}_{iI}, \nonumber \\
{\boldsymbol \xi}_i^{ee}({\bf R})&=\sum_{j\ne i} \eta(r_{ij}) {\bf r}_{ij}, \nonumber  \\
{\boldsymbol \xi}_i^{een}({\bf R})&=\sum_I \sum_{j\ne i} [\theta_1(r_{ij},r_{iI},r_{jI}) {\bf r}_{ij} + \theta_2(r_{ij},r_{iI},r_{jI}) {\bf r}_{iI}],
\end{align}
where ${\bf r}_{ij}={\bf r}_i-{\bf r}_j$, ${\bf r}_{iI}={\bf r}_i-{\bf r}_I$ and we sum over all nuclei $I$ and electrons $j$. 
The $\mu$, $\eta$ and $\theta_1$ with $\theta_2$ terms are similar to one, two and three-body Jastrow terms present in $U_{corr}$ of trial wave function, Eq.~(\ref{eq:slater-jastrowwf}),
and are further expanded as 
\begin{align}
\label{eq:bffunctions}
\mu(r)&=\sum_k c_k a_k(r), \nonumber  \\
\eta(r)&=\sum_k d_k b_k(r), \nonumber  \\
\theta_{1(2)}(r_{ij},r_{iI},r_{jI})&=\sum_{klm} g_{klm}^{1(2)} a_k(r_{iI})a_l(r_{jI})b_m(r_{ij}). 
\end{align}

\section{Backflow Wave function Results}
In this section, we present VMC and DMC results obtained with above implementation of backflow 
correlations for determinant and pfaffian wave functions. The Jastrow factors and pseudopotentials 
are identical to ones used in Sec.~\ref{sec:4}. The distance-dependent basis functions 
$\{a\}$ and $\{b\}$  used in the Eq.~(\ref{eq:bffunctions}) are chosen  either as Gaussians 
centered on the nucleus or polynomial Pad\' e functions~\cite{qwalk} to preserve 
the electron-electron and electron-nucleus 
(when used with pseudopotentials) cusp conditions~\cite{Kato57,Pack66}.
The sets of variational parameters $\{c\}$, $\{d\}$ and $\{g\}$ are minimized in the similar fashion as in Sec.~\ref{sec:4}
with respect to energy or mixture of energy and variance~\cite{cyrus2}. 
In addition, all electron-electron coefficients ($\{d_k\}$ and $\{g_{klm}^{1(2)}\}$ with fixed $k$ and $l$) are 
allowed to be different for spin-like and for spin-unlike electron pairs.

\subsection{Homogeneous electron gas}
We benchmark our implementation of the backflow correlations on the homogeneous electron gas (HEG). 
The HEG system of 54 unpolarized electrons in the simple cubic simulation cell with periodic boundary conditions 
was studied  before~\cite{kwon4,markus,rios_bf}. 
We use the backflow displacement, Eq.~(\ref{eg:bfdisplacement}), with only ${\boldsymbol \xi}_i^{ee}$ being non-zero and 
let $\eta(r)$ in Eq.~(\ref{eg:bfterms}) to be different for spin-like and for spin-unlike electron pairs. 
These functions are further expanded in the basis of polynomial Pad\' e functions~\cite{qwalk} with cutoff 
equal to half of the simulation cell.  

We compare our results for the following three densities of $r_s=1$, $5$ and $20$ (see Table~\ref{table:heg}). 
First, it is clear that the HF and Slater--Jastrow (SJ) fixed-node DMC energies are in good agreement with  previous results~\cite{kwon4,markus,rios_bf}.
Second, due to the omission of the three-body correlations from Jastrow factor and also from backflow displacement, 
it is expected that we obtain higher VMC energies and variances for SJ and backflow displaced SJ (SJBF) trial wave functions. Nevertheless, 
our fixed-node DMC energies for SJBF trial wave functions closely match the results of Kwon {\it et al.\/}\cite{kwon4}, and 
only slightly deviate at higher densities from results od Rios {\it et al.\/}\cite{rios_bf}
 
\begin{table}
\caption{VMC and fixed-node DMC energies per electron and variances of local energies for various 
trial wave functions (S, Slater; SJ, Slater--Jastrow; SJBF, backflow correlated SJ) for 3D unpolarized HEG of 54 electrons.}
\begin{ruledtabular}
\begin{tabular}{l l l c c }
\multicolumn{1}{l}{$r_s$} & \multicolumn{1}{l}{Method} & \multicolumn{1}{l}{WF}& \multicolumn{1}{c}{$E/N$[a.u./electron]} & \multicolumn{1}{c}{$\sigma^2$[a.u.$^2$]}\\
\hline
1.0 & HF  & S    &  0.56925(2)  &  19.3(1)   \\
    & VMC & SJ   &  0.53360(4)  &  1.26(4)   \\
    &     & SJBF &  0.53139(4)  &  0.81(4)   \\
    & DMC & SJ   &  0.53087(4)  &  -        \\ 
    &     & SJBF &  0.52990(4)  &  -        \\ 
\hline            
5.0 & HF  & S    &  -0.056297(7) &  0.776(4)  \\
    & VMC & SJ   &  -0.075941(6) &  0.0636(1) \\
    &     & SJBF &  -0.078087(4) &  0.0236(1) \\
    & DMC & SJ   &  -0.07862(1)  &  -         \\ 
    &     & SJBF &  -0.07886(1)  &  -         \\    
\hline            
20.0 & HF  & S   &  -0.022051(2) &  0.0482(1)  \\
     & VMC & SJ  &  -0.031692(2) &  0.000711(4) \\
     &     & SJBF&  -0.031862(1) &  0.000437(1) \\
     & DMC & SJ  &  -0.031948(2) &  -        \\ 
     &     & SJBF&  -0.032007(2) &  -        \\    
\end{tabular}
\end{ruledtabular}
\label{table:heg}
\end{table}

\subsection{Carbon atom and dimer}
The backflow correlations in single-determinant Slater--Jastrow trial wave functions were 
recently applied also to inhomogeneous systems~\cite{drummond_bf,rios_bf,gurtubay_bf}. 
They were demonstrated to capture additional few percent of the correlation energy 
but being somewhat shy of the goal of more than 99\%, with the only exception 
of Li atom.  
It was also suggested that the backflow by itself is unlikely to change the number of nodal cells. 
These observations let to further studies of  backflow combined with the wave functions that  
have the minimal number of nodal cells --- an important topological property associated 
with ground state wave functions\cite{davidnode,dariobe,lubos_nodeprl,lubos_nodeprb}.
One of the successes of this scheme are the very recent results of Brown {\it et al.\/}\cite{brown_bf}  
obtained from
backflow correlated CI--Jastrow  wave functions applied to first row  all-electron atoms.
In this study, we further test the limits of the backflow correlations to decrease 
the fixed node errors of the CI--Jastrow wave functions and, for the first time,
also include backflow into the pfaffian--Jastrow pairing wave functions. 
Below is a brief discussion of our implementation and results for carbon atom and dimer systems.

\begin{figure}
\begin{center}
\includegraphics[width=\columnwidth]{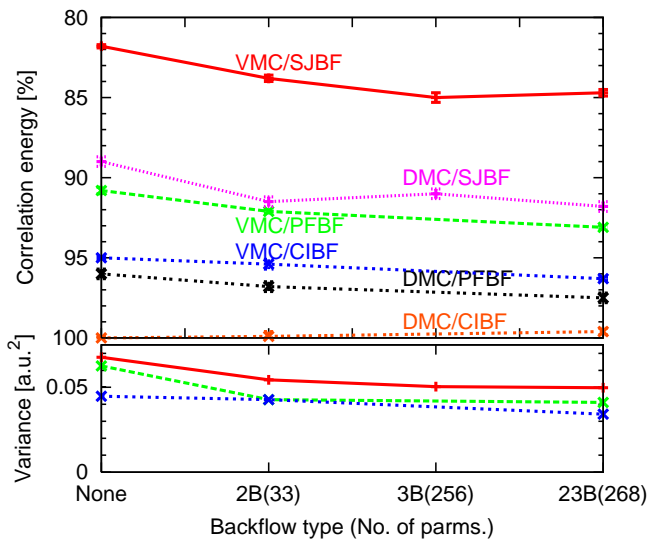}
\end{center}
\caption{
(color online) Upper figure: Percentages of correlation energy  from VMC and DMC methods versus a number of backflow parameters
with backflow correlated Slater--Jastrow (SJBF), pfaffian--Jastrow (PFBF)  and CI--Jastrow (CIBF) trial wave functions 
for C atom (2B, electron-nucleus and electron-electron terms; 3B, all electron-electron-nucleus terms; 
23B for all terms together). Lines connecting the points serve only as a guide to the eye.
Lower figure: Variance of the local energy versus a number of backflow parameters.}

\label{fig:C_bf}
\end{figure}

\begin{figure}
\begin{center}
\includegraphics[width=\columnwidth]{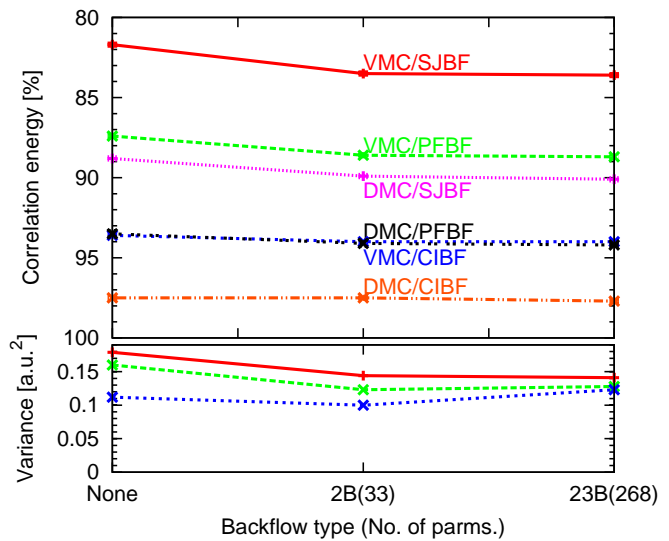}
\end{center}
\caption{(color online) Same as Fig.~\ref{fig:C_bf} but for C dimer.}
\label{fig:C2_bf}
\end{figure}

In the inhomogeneous backflow, each electron's 
coordinate is correlated by the displacement as given by Eq.~(\ref{eg:bfdisplacement}), 
while the functions $\eta$ and $\theta_{1,2}$ are allowed to be spin-dependent.
We use up to 11 Gaussian basis functions to fit the  $\mu$ and $\eta$ functions, while the three-body 
functions $\theta_{1,2}$ are limited to a product of $4\times4\times4$ Gaussians.
The main results are plotted in Figs.~\ref{fig:C_bf}~and~\ref{fig:C2_bf} and
detailed numerical results are summarized  
in Tables~\ref{table:bf:C} and~\ref{table:bf:C2} of Appendix~\ref{appendix:table}. 
The backflow correlations are able to capture additional few percent of the correlation energy 
for both Slater--Jastrow and pfaffian--Jastrow wave functions. 
Another important feature of backflow is 20-30\% decrase in variances of local energy with respect 
to the wave functions without backflow correlations. We find that for the fully optimized backflow, 
the spin-unlike electron-electron functions are almost order of magnitude larger than spin-like ones as well as electron-nucleus functions.
The gains are systematic with increasing number of parameters, however we do not 
find the three-body terms as important as reported in previous study~\cite{rios_bf}. 
This difference can be attributed to two main reasons---we use a different basis to expand the three-body 
functions $\theta_{1,2}$ and we also eliminate atomic cores by pseudopotentials. It is plausible
that for systems with core electrons the three-body correlations are more important due to the 
strong variations of orbitals close to the nucleus.

Finally, let us discuss the difference between the two systems with respect to missing correlation energy. 
For the C atom, we have shown 
previously~\cite{pfaffianprl} that less than 100 determinants gives more than $99\%$ of correlation energy ($E_{corr}$). 
The C dimer's fixed node errors are more pronounced, since the 148 determinants with re-optimized weights 
give only 97.5(1)\% in a close agreement
with recent calculations by Umrigar {\it et al.\/}\cite{umrigarC2}.
Employing backflow correlations for our 148 determinant CI-Jastrow wave function gives no apparent gain in $E_{corr}$ 
except for decrease in the variance of local energy.  
The improvement for the pfaffian-Jastrow wave function is also very modest (less than 1\%).
Our results suggest that to reach beyond 99\% of correlation one still needs 
complicated multi-reference wave functions, even after including quite general forms
of the backflow correlations. 

\section{\label{sec:level5} Conclusions}

To summarize, we have proposed pfaffians with
singlet pair, triplet pair and unpaired
orbitals as variationally rich and compact wave functions.
They offer significant and systematic improvements
over commonly used Slater determinant-based wave functions. 
We have included a set of key mathematical 
identities with proofs, which are needed 
for the evaluation and update of the pfaffians.
We have also shown connections of HF and BCS wave functions to more general 
pfaffian wave function.
Further, we have explored multi-pfaffian wave functions,
which enabled us to capture additional correlation.
While for atomic systems the results are comparable to large-scale CI wave functions\cite{pfaffianprl}, 
equivalent accuracy in molecular systems most probably require much larger multi-pfaffian expansions
than we have explored. As another test of the variational potential of
pairing, we have employed the fully-antisymmetrized independent pairs wave function 
in pfaffian form and we have found that it does not lead to additional gains 
in correlation energy. 
We therefore conclude that more general functional forms 
together with more robust large-scale optimization methods might be necessary 
in order to obtain further improvements. 
The gains in correlation energy for pfaffians come from
improved fermion nodes which are significantly closer to the exact ones than the HF nodes and exhibit 
the correct topology with the minimal number of two nodal cells.

In the second part of the paper, we have presented the application of pfaffian and multi-determinantal wave functions 
with backflow correlations to chemical systems. Results for two testing cases of C atom and its dimer 
show promising gains in correlation energies, decreases in variances and improvements in the nodal structure.
Our results also indicate that 
 accurate description of molecular systems with fixed-node errors bellow $1\%$ of correlation energy
requires optimized multi-reference wave functions and inclusion of backflow then appears less
favorable considering significant computational cost, especially for evaluation  of the 
nonlocal pseudopotential operators.  

\begin{acknowledgments}
We gratefully acknowledge the support by 
NSF Grants No. DMR-0121361, PHY-0456609, and  EAR-0530110 
and the computer time at PAMS NCSU and NCSA facilities as well the INCITE allocation at ORNL.
\end{acknowledgments}

\appendix
\section{Proof of Cayley's identity}\label{appendix:Cayley}
In order to prove the statement in Eq.~(\ref{eq:cayley}) we will proceed by induction. For $n=2$ it is true that
\begin{equation*}
{\rm det} \left[\begin{array}{cc}
0  & b_{12} \\
-a_{12}  & 0 \end{array}\right]
={\rm pf} \left[\begin{array}{cc}
0  & b_{12}\\
-b_{12} & 0\end{array}\right]
{\rm pf} \left[\begin{array}{cc}
0  & a_{12} \\
-a_{12} & 0\end{array}\right].
\end{equation*}
For even $n$ greater than $2$, determinant of the matrix of interest 
can be expanded through its cofactors as
\begin{align}\label{eq:a:expand}
 {\rm det}& \left[\begin{array}{ccccc}
0  & b_{12}  & b_{13} &\ldots &  b_{1,n}\\
-a_{12}  & 0  & a_{23} & \ldots &  a_{2,n}\\
-a_{13}  & -a_{23} & 0  & \ldots &  a_{3,n}\\
 \vdots & \vdots & \vdots &  \ddots &  \vdots \\
-a_{1,n} & -a_{2,n} & -a_{3,n} & \ldots &  0\\
\end{array}\right]=\sum_k -a_{1,k} C(k,1) \nonumber \\ 
&=\sum_k  \sum_l  -a_{1,k} b_{1,l} C(k,1;1,l).
\end{align}
The cofactor can be written as
\begin{align}
C(k,1;1,l)=(-1)^{k+l+1}{\rm det} \left[A(k,1;1,l)\right],
\end{align}
where the cofactor matrix  is given by
\begin{align}
\small
A(k,1;1,l)=
\left[
\begin{array}{ccccccc}
0 & a_{23}  & \ldots & a_{2,k} & \ldots &  a_{2,n}\\
-a_{23}  & 0  & \ldots &  a_{3,k} & \ldots & a_{3,n}\\
 \vdots & \vdots & \ddots & \vdots & \ddots & \vdots  \\
-a_{2,l} & -a_{3,l} & \ldots & -a_{k,l} &\ldots & a_{l,n}\\
\vdots & \vdots & \ddots & \vdots & \ddots & \vdots  \\
-a_{2,n} & -a_{3,n} & \ldots & -a_{k,n} &\ldots & 0
\end{array}\right].
\end{align}
At this point we would like to use an induction step and rewrite
the determinant cofactor as a product of two pfaffians [Cayley's identity Eq.~(\ref{eq:cayley})].
This would allow us to demonstrate that the expansion is identical to 
the expansion of pfaffians in minors.
In order to do so, however, we have
to shift the $k$-th column by pair column exchanges, so it becomes
the {\em last} column and, similarly, we have to shift the $l$-th row by
pair exchanges, so it becomes the last row.
This involves $k$ pair exchanges of columns and $l$ pair exchanges or rows and 
can be represented by unitary matrices $U_k$ and $U_l$.
It is necessary to invoke these operations so that the 
matrix gets into a form directly amenable for the Cayley's identity, i.e.,
the matrix has to be in a manifestly skew-symmetric form. 
(The sign change from the row/columns exchanges will prove irrelevant as we will 
show below.)
The transformed matrix is given by 
\begin{align}
A'(k,1;1,l)=U_kA(k,1;1,l)U_l
\end{align}
and has all zeros on the diagonal with the exception of the last element which is equal to $-a_{k,l}$. 
The last row is given by
\begin{align} 
{\bf v}_r=&(-a_{2,l},\ldots, -a_{k-1,l},-a_{k+1,l},\ldots \nonumber \\
&\ldots,-a_{l-1,l},a_{l,l+1},\ldots, a_{l,n},-a_{k,l}), 
\end{align}
while the last column is given as following
\begin{align}
{\bf v}_c^T=&(a_{2,k},\ldots,a_{k-1,k},-a_{k,k+1},\ldots \nonumber \\
&\ldots,-a_{k,l-1},-a_{k,l+1},\ldots,-a_{k,n},-a_{k,l})^T.
\end{align}
The only non-zero diagonal element $-a_{k,l}$ can be eliminated, 
once we realize that its cofactor contains a determinant of
a skew-symmetric matrix of odd degree
which always vanishes (proof by Jacobi\cite{Jacobi}).

Now we are ready to perform the induction step, namely to use the 
property that the determinant of a $2(n-1)\times 2(n-1)$ matrix
can be written as given by the Cayley's identity Eq.~(\ref{eq:cayley}).
We obtain
\begin{align}
{\rm det}[U_kA(k,1;1,l)U_l]&={\rm det}[A'(k,1;1,l)] \\
 &={\rm pf}[A'(1,k;1,k)]\,{\rm pf}[A'(1,l;1,l)]. \nonumber 
\end{align}
We can now apply the inverse unitary transformations and
shift back the columns (and by the skew-symmetry the corresponding rows) in the
first pfaffian and, similarly, the rows (and corresponding columns) 
in the second. This enables us to write
\begin{align}
{\rm pf}&[A'(1,k;1,k)]\,{\rm pf}[A'(1,l;1,l)]\nonumber \\
&={\rm pf}[U_l^{-1}A(1,k;1,k)U_l]\,{\rm pf}[U_kA(1,l;1,l)U_k^{-1}]\nonumber \\
&={\rm pf}[A(1,k;1,k)]\,{\rm pf}[A(1,l;1,l)],
\end{align}
where we have used the identity given by Eq.~(\ref{eq:pfident4}).
We can therefore finally write
\begin{align}
C(k,1;1,l)&=(-1)^{k+l+1}{\rm pf}[A(1,k;1,k)]{\rm pf}[A(1,l;1,l)] \nonumber \\
&=-P_c(a_{1,k})P_c(a_{1,l}),
\end{align}
where $P_c$ denotes a pfaffian cofactor as defined in (\ref{eq:pfcof}). 
Therefore, the determinant expansion in Eq.~(\ref{eq:a:expand}) equals to
\begin{align}
\sum_{k,l} -a_{1,k} b_{1,l} C(k,1;1,l)&=\sum_{k,l} a_{1,k} b_{1,l} P_c(a_{1,k})P_c(a_{1,l}) \nonumber \\
&={\rm pf}[A]{\rm pf}[B]
\end{align}
with matrices $A$ and $B$ defined as in Eq.~(\ref{eg:inverseupdate}).
This concludes the proof of the more general form of the Cayley's identity. Note, if $B=A$,
we trivially obtain well-known formula for the square of
pfaffian [Eq.~(\ref{eq:pfident2})].

\section{}\label{appendix:table}
\begingroup
\begin{table*}
\caption{ Total energies for C, N and O atoms and their dimers with 
amounts of the correlation energy recovered in VMC and DMC methods with wave functions 
as discussed in the text. Unless noted otherwise, 
the numbers in parentheses are the statistical errors in the last digit from corresponding QMC calculation.
Energies are in Hartree atomic units.
For C, N, O atoms we used the correlation energies by Dolg \cite{dolgcpl}(0.1031, 0.1303, 0.1937 a.u.).
For the estimation of correlation energies of dimers we needed accurate HF energies at experimental distances\cite{NIST-JANAF}
and the estimated exact total energies. 
Each exact total energy was estimated as a sum of total energies of constituent atoms minus experimental binding energy\cite{NIST-JANAF, UBJ, HH}
adjusted for experimental zero-point energy\cite{HH}.} 
\begin{ruledtabular}
\begin{tabular}{l l d d d d d d }
 \multicolumn{1}{l}{Method}&\multicolumn{1}{l}{WF}&\multicolumn{1}{c}{C}&\multicolumn{1}{c}{E$_{corr}$[\%]} & \multicolumn{1}{c}{N} &\multicolumn{1}{c}{E$_{corr}$[\%]}&\multicolumn{1}{c}{O}& \multicolumn{1}{c}{E$_{corr}$[\%]}\\
\hline
HF  & S                & -5.31471   &   0     & -9.62892   & 0       & -15.65851   &  0      \\
VMC & SJ            & -5.3939(4) & 76.8(4) & -9.7375(1) & 83.3(1) & -15.8210(6) & 83.9(3) \\
    & BCS           & -5.4061(2) & 88.6(2) & -9.7427(3) & 87.3(2) & -15.8250(3) & 86.0(2) \\
    & STU           & -5.4068(2) & 89.3(2) & -9.7433(1) & 87.8(1) & -15.8255(3) & 86.2(2)  \\
DMC & SJ            & -5.4061(3) & 88.6(2) & -9.7496(2) & 92.6(2) & -15.8421(2) & 94.8(1) \\
    & BCS           & -5.4140(2) & 96.3(2) & -9.7536(2) & 95.7(2) & -15.8439(4) & 95.7(2) \\
    & STU           & -5.4139(2) & 96.2(2) & -9.7551(2) & 96.8(1) & -15.8433(3) & 95.4(2)  \\
Est.& Exact        & -5.417806  & 100     & -9.759215  & 100     & -15.85216   & 100  \\
\hline
\multicolumn{1}{l}{Method}& \multicolumn{1}{l}{WF} &  \multicolumn{1}{c}{C$_2$} &  \multicolumn{1}{c}{E$_{corr}$[\%]} &  \multicolumn{1}{c}{N$_2$} &  \multicolumn{1}{c}{E$_{corr}$[\%]} &  \multicolumn{1}{c}{O$_2$} &  \multicolumn{1}{c}{E$_{corr}$[\%]} \\
\hline
HF  & S               & -10.6604    & 0       & -19.4504   &  0       & -31.3580    & 0 \\
VMC & SJ            & -10.9579(4) & 72.9(1) & -19.7958(5)&  80.0(1) & -31.7858(6) & 79.6(1)\\
    & BCS           & -11.0059(4) & 84.7(1) & -19.8179(6)&  85.0(1) & -31.8237(4) & 86.7(1)\\
    & STU           & -11.0062(3) & 84.8(1) & -19.821(1) &  85.8(2) & -31.8234(4) & 86.6(1)\\
DMC & HF            & -11.0153(4) & 87.0(1) & -19.8521(3)&  93.0(1) & -31.8649(5) & 94.3(1)\\
    & BCS           & -11.0416(3) & 93.5(1) & -19.8605(6)&  94.9(1) & -31.8664(5) & 94.6(1)\\
    & STU           & -11.0421(5) & 93.6(1) & -19.8607(4)&  95.0(1) & -31.8654(5) & 94.4(1)\\
Est.& Exact\footnotemark[3]  & -11.068(5)\footnotemark[1]  & 100.0(10)  & -19.8825(6)\footnotemark[2]   &  100.0(1)     & -31.8954(1)\footnotemark[2]    & 100.0(1)    \\
\end{tabular}
\end{ruledtabular}
\footnotetext[1] { 
There is rather large discrepancy in the experimental values of C$_2$ binding energy
($141.8(9)$\cite{NIST-JANAF}, $143(3)$\cite{HH} and $145.2(5)$ kcal/mol\cite{UBJ}). 
For the estimation of exact energy we have taken the average value of $143(3)$ kcal/mol.}
\footnotetext[2] {Experimental binding energies taken from ref. [\onlinecite{NIST-JANAF}].}
\footnotetext[3] {The error bars on estimated exact total energies are due to experiment.}
\label{energies:1}
\end{table*}
\endgroup

\begin{table*}
\caption{VMC and DMC energies and variances of local energy for Slater--Jastrow (SJ), pfaffian--Jastrow (PF) and CI-Jastrow (CI) trial wave functions with backflow (BF) correlations 
for C atom. Notation is the same as in Fig.~\ref{fig:C_bf}.}
\begin{ruledtabular}
\begin{tabular}{l c c c c c  c c c c c  }
\multicolumn{1}{l}{Method} &\multicolumn{1}{c}{WF} & \multicolumn{1}{c}{$N_\mu$}  & \multicolumn{1}{c}{$N_\eta$} & \multicolumn{1}{c}{$N_{\theta_1}$}  & \multicolumn{1}{c}{$N_{\theta_2}$} &
 \multicolumn{1}{c}{N$_p$} & \multicolumn{1}{c}{E [a.u.]} &  \multicolumn{1}{c}{$\sigma^2$ [a.u.$^2$]} & \multicolumn{1}{c}{E$_{corr}$[\%]} \\
\hline 
HF  & S & -& - & - & -& -& -5.31471 & - & 0.0 \\
\hline 
VMC & SJ      & -  & -  & -   & -   & -   & -5.3990(1) & 0.0677 & 81.8(1)\\
    & SJBF2B  & 11 & 22 & -   & -   & 33  & -5.4011(2) & 0.0544 & 83.8(2)\\ 
    & SJBF3B  & -  & -  & 128 & 128 & 256 & -5.4023(3) & 0.0504 & 85.0(3)\\
    & SJBF23B & 4  & 8  & 128 & 128 & 268 & -5.4020(2) & 0.0498 & 84.7(2)\\
    & PF      & -  & -  & -   & -   & -   & -5.4083(2) & 0.0626 & 90.8(2)\\
    & PFBF2B  & 11 & 22 & -   & -   & 33  & -5.4097(1) & 0.0427 & 92.1(1)\\
    & PFBF23B & 4  & 8  & 128 & 128 & 268 & -5.4107(1) & 0.0411 & 93.1(1)\\
    & CI\footnotemark[1]      & -  & -  & -   & -   & -   & -5.4127(1) & 0.0447 & 95.0(1)\\
    & CIBF2B  & 11 & 22 & -   & -   & 33  & -5.4131(3) & 0.0427 & 95.4(3)\\
    & CIBF23B & 4  & 8  & 128 & 128 & 268 & -5.4140(1) & 0.0342 & 96.3(1)\\
\hline 
DMC & SJ      & -  & -  & -   & -   & -   & -5.4065(3) & - & 89.0(3)\\
    & SJBF2B  & 11 & 22 & -   & -   & 33  & -5.4090(3) & - & 91.5(3)\\
    & SJBF3B  & -  & -  & 128 & 128 & 256 & -5.4085(3) & - & 91.0(3)\\
    & SJBF23B & 4  & 8  & 128 & 128 & 268 & -5.4094(3) & - & 91.8(3)\\
    & PF      & -  & -  & -   & -   & -   & -5.4137(3) & - & 96.0(3)\\
    & PFBF2B  & 11 & 22 & -   & -   & 33  & -5.4145(3) & - & 96.8(3)\\
    & PFBF23B & 4  & 8  & 128 & 128 & 268 & -5.4152(3) & - & 97.5(3)\\
    & CI      & -  & -  & -   & -   & -   & -5.4178(1) & - & 100.0(1)\\
    & CIBF2B  & 11 & 22 & -   & -   & 33  & -5.4177(3) & - & 99.9(3) \\
    & CIBF23B & 4  & 8  & 128 & 128 & 268 & -5.4174(2) & - & 99.6(2) \\
\hline
Est.& Exact & - & -& - & - & - & -5.417806 & - & 100.0 \\
\end{tabular}
\footnotetext[1] {Wave function consists of 100 determinants re-optimized in VMC.}
\end{ruledtabular}
\label{table:bf:C}
\end{table*}

\begin{table*}
\caption{Slater--Jastrow (SJ), pfaffian--Jastrow (PF)  and CI--Jastrow (CI) wave functions with backflow (BF) correlations for C dimer.
Notation is the same as in Fig.~\ref{fig:C_bf}.}
\renewcommand{\thefootnote}{\alph{footnote}}
\renewcommand{\thempfootnote}{\alph{mpfootnote}}
\centering
\begin{ruledtabular}
\begin{tabular}{l c c c c c  c c c c c  }
\multicolumn{1}{l}{Method} &\multicolumn{1}{c}{WF} & \multicolumn{1}{c}{$N_\mu$}  & \multicolumn{1}{c}{$N_\eta$} & \multicolumn{1}{c}{$N_{\theta_1}$}  & \multicolumn{1}{c}{$N_{\theta_2}$} &
 \multicolumn{1}{c}{N$_p$} & \multicolumn{1}{c}{E [a.u.]} &  \multicolumn{1}{c}{$\sigma^2$ [a.u.$^2$]} & \multicolumn{1}{c}{E$_{corr}$[\%]} \\
\hline 
HF & S & -& - & - & -& -& -10.6604 & - & 0.0 \\
\hline 
VMC & SJ\footnotemark[1]     & -  & -  & -   & -   & -   & -10.9936(4) & 0.179 & 81.7(1)\\
    & SJBF2B                 & 11 & 22 & -   & -   & 33  & -11.0012(3) & 0.144 & 83.5(1)\\
    & SJBF23B                & 4  & 8  & 128 & 128 & 268 & -11.0014(2) & 0.141 & 83.6(1)\\
    & PF\footnotemark[2]     & -  & -  & -   & -   & -   & -11.0171(2) & 0.160 & 87.4(1)\\
    & PFBF2B                 & 11 & 22 & -   & -   & 33  & -11.0223(3) & 0.123 & 88.7(1)\\
    & PFBF23B                & 4  & 8  & 128 & 128 & 268 & -11.0223(2) & 0.128 & 88.7(1)\\
    & CI\footnotemark[3]     &  - & -  & -   & -   & -   & -11.0420(4) & 0.112 & 93.6(1)\\
    & CIBF2B                 & 11 & 22 & -   & -   & 33  & -11.0440(3) & 0.100 & 94.0(1)\\
    & CIBF23B                & 4  & 8  & 128 & 128 & 268 & -11.0438(3) & 0.123 & 94.0(1)\\
\hline 
DMC & SJ      & -  & -  & -   & -   & -   & -11.0227(2) & - & 88.8(1)\\
    & SJBF2B  & 11 & 22 & -   & -   & 33  & -11.0269(4) & - & 89.9(1)\\
    & SJBF23B & 4  & 8  & 128 & 128 & 268 & -11.0280(3) & - & 90.1(1)\\
    & PF      & -  & -  & -   & -   & -   & -11.0419(9) & - & 93.5(2)\\
    & PFBF2B  & 11 & 22 & -   & -   & 33  & -11.0443(6) & - & 94.1(2)\\
    & PFBF23B & 4  & 8  & 128 & 128 & 268 & -11.0447(3) & - & 94.2(1)\\
    & CI      &  - & -  & -   & -   & -   & -11.0579(5) & - & 97.5(1)\\
    & CIBF2B  & 11 & 22 & -   & -   & 33  & -11.0580(4) & - & 97.5(1)\\
    & CIBF23B & 4  & 8  & 128 & 128 & 268 & -11.0585(5) & - & 97.7(1)\\
\hline
Est.& Exact & - & -& - & - & - & -11.068(5) & - & 100.0 \\
\end{tabular}
\footnotetext[1] {Slater determinant contains PBE DFT orbitals.}
\footnotetext[2] {Same PBE DFT orbitals are used also in PF wave function.}
\footnotetext[3] {Uses natural orbitals with weights of the 148 determinants re-optimized in VMC.}
\end{ruledtabular}
\label{table:bf:C2}
\end{table*}

\bibliography{thesismb}

\end{document}